\documentclass[11pt]{article}

\catcode`\@=11
\long\def\@makefntext#1{
\protect\noindent \hbox to 3.2pt {\hskip-.9pt
$^{{\eightrm\@thefnmark}}$\hfil}#1\hfill}       

\def\@makefnmark{\hbox to 0pt{$^{\@thefnmark}$\hss}}    

\def\ps@myheadings{\let\@mkboth\@gobbletwo
\def\@oddhead{\hbox{}
\rightmark\hfil\eightrm\thepage}
\def\@oddfoot{}\def\@evenhead{\eightrm\thepage\hfil
\leftmark\hbox{}}\def\@evenfoot{}
\def\sectionmark##1{}\def\subsectionmark##1{}}

\newcounter{sectionc}\newcounter{subsectionc}\newcounter{subsubsectionc}
\renewcommand{\section}[1] {\vspace{12pt}\addtocounter{sectionc}{1}
\setcounter{subsectionc}{0}\setcounter{subsubsectionc}{0}\noindent
    \par\vspace{5pt}}
\renewcommand{\subsection}[1] {\vspace{12pt}\addtocounter{subsectionc}{1}
    \setcounter{subsubsectionc}{0}\noindent
    {\bf\thesectionc.\thesubsectionc. {\kern1pt \bfit #1}}\par\vspace{5pt}}
\renewcommand{\subsubsection}[1] {\vspace{12pt}\addtocounter{subsubsectionc}{1}
    \noindent{\tenrm\thesectionc.\thesubsectionc.\thesubsubsectionc.
    {\kern1pt \tenit #1}}\par\vspace{5pt}}

\newcounter{appendixc}
\newcounter{subappendixc}[appendixc]
\newcounter{subsubappendixc}[subappendixc]
\renewcommand{\thesubappendixc}{\Alph{appendixc}.\arabic{subappendixc}}
\renewcommand{\thesubsubappendixc}
        {\Alph{appendixc}.\arabic{subappendixc}.\arabic{subsubappendixc}}

\renewcommand{\appendix}[1] {\vspace{12pt}
        \refstepcounter{appendixc}
        \setcounter{figure}{0}
        \setcounter{table}{0}
        \setcounter{lemma}{0}
        \setcounter{theorem}{0}
        \setcounter{corollary}{0}
        \setcounter{definition}{0}
        \setcounter{equation}{0}
        \renewcommand{\thefigure}{\Alph{appendixc}.\arabic{figure}}
        \renewcommand{\thetable}{\Alph{appendixc}.\arabic{table}}
        \renewcommand{\theappendixc}{\Alph{appendixc}}
        \renewcommand{\thelemma}{\Alph{appendixc}.\arabic{lemma}}
        \renewcommand{\thetheorem}{\Alph{appendixc}.\arabic{theorem}}
        \renewcommand{\thedefinition}{\Alph{appendixc}.\arabic{definition}}
        \renewcommand{\thecorollary}{\Alph{appendixc}.\arabic{corollary}}
        \renewcommand{\theequation}{\Alph{appendixc}.\arabic{equation}}
   \noindent{\tenbf Appendix \theappendixc. #1}\par\vspace{5pt}}
\newcommand{\subappendix}[1] {\vspace{12pt}
        \refstepcounter{subappendixc}
        \noindent{\bf Appendix \thesubappendixc. {\kern1pt \bfit #1}}
        \par\vspace{5pt}}
\newcommand{\subsubappendix}[1] {\vspace{12pt}
        \refstepcounter{subsubappendixc}
        \noindent{\rm Appendix \thesubsubappendixc. {\kern1pt \tenit #1}}
        \par\vspace{5pt}}

\newcommand{\textlineskip}{\baselineskip=13pt}
\newcommand{\smalllineskip}{\baselineskip=10pt}

\def\eightcirc{
\begin{picture}(0,0)
\put(4.4,1.8){\circle{6.5}}
\end{picture}}
\def\eightcopyright{\eightcirc\kern2.7pt\hbox{\eightrm c}}

\def\abstracts#1#2#3{{
    \centering{\begin{minipage}{4.5in}\baselineskip=10pt\footnotesize
    \parindent=0pt #1\par
    \parindent=15pt #2\par
    \parindent=15pt #3
    \end{minipage}}\par}}

\renewenvironment{thebibliography}[1]
    {\frenchspacing
     \ninerm\baselineskip=11pt
     \begin{list}{\arabic{enumi}.}
        {\usecounter{enumi}\setlength{\parsep}{0pt}
     \setlength{\leftmargin 12.7pt}{\rightmargin 0pt} 
    \setlength{\leftmargin 17pt}{\rightmargin 0pt}   
    \setlength{\leftmargin 22pt}{\rightmargin 0pt}   
         \setlength{\itemsep}{0pt} \settowidth
    {\labelwidth}{#1.}\sloppy}}{\end{list}}

\newcounter{itemlistc}
\newcounter{romanlistc}
\newcounter{alphlistc}
\newcounter{arabiclistc}

\newcommand{\fcaption}[1]{
        \refstepcounter{figure}
        \setbox\@tempboxa = \hbox{\footnotesize Fig.~\thefigure. #1}
        \ifdim \wd\@tempboxa > 5in
           {\begin{center}
        \parbox{5in}{\footnotesize\smalllineskip Fig.~\thefigure. #1}
            \end{center}}
        \else
             {\begin{center}
             {\footnotesize Fig.~\thefigure. #1}
              \end{center}}
        \fi}

\def\@citex[#1]#2{\if@filesw\immediate\write\@auxout
    {\string\citation{#2}}\fi
\def\@citea{}\@cite{\@for\@citeb:=#2\do
    {\@citea\def\@citea{,}\@ifundefined
    {b@\@citeb}{{\bf ?}\@warning
    {Citation `\@citeb' on page \thepage \space undefined}}
    {\csname b@\@citeb\endcsname}}}{#1}}

\newif\if@cghi
\def\citelow{\@cghifalse\@ifnextchar [{\@tempswatrue
    \@citex}{\@tempswafalse\@citex[]}}

\def\@refcitex[#1]#2{\if@filesw\immediate\write\@auxout
    {\string\citation{#2}}\fi
\def\@citea{}\@refcite{\@for\@citeb:=#2\do
    {\@citea\def\@citea{, }\@ifundefined
    {b@\@citeb}{{\bf ?}\@warning
    {Citation `\@citeb' on page \thepage \space undefined}}
    \hbox{\csname b@\@citeb\endcsname}}}{#1}}

\def\@refcite#1#2{{#1\if@tempswa\typeout
        {IJCGA warning: optional citation argument
    ignored: `#2'} \fi}}

\def\refcite{\@ifnextchar[{\@tempswatrue
    \@refcitex}{\@tempswafalse\@refcitex[]}}

\def\pmb#1{\setbox0=\hbox{#1}
    \kern-.025em\copy0\kern-\wd0
    \kern.05em\copy0\kern-\wd0
    \kern-.025em\raise.0433em\box0}

\def\fnt#1#2{\footnotetext{\kern-.3em
    {$^{\mbox{\scriptsize #1}}$}{#2}}}

\def\runninghead#1#2{\pagestyle{myheadings}
\markboth{{\protect\footnotesize\it{\quad #1}}\hfill}
{\hfill{\protect\footnotesize\it{#2\quad}}}}
\font\tenrm=cmr10
\font\tenit=cmti10
\font\tenbf=cmbx10
\font\bfit=cmbxti10 at 10pt
\font\ninerm=cmr9

\font\eightrm=cmr8

\def\qed{\hbox{${\vcenter{\vbox{            
   \hrule height 0.4pt\hbox{\vrule width 0.4pt height 6pt
   \kern5pt\vrule width 0.4pt}\hrule height 0.4pt}}}$}}

\usepackage{graphicx}
\usepackage{dcolumn}
\usepackage{bm}

\begin{document}

\newpage

\runninghead{Perales {\em et al}} {Synchronization of Frenet-Serret with Malasoma}

\normalsize\textlineskip
\thispagestyle{empty}
\setcounter{page}{1}


\vspace*{0.88truein}

\centerline{\bf Int. J. Theor. Phys., accepted}
\bigskip
\centerline{\bf SYNCHRONIZATION OF THE FRENET-SERRET LINEAR SYSTEM} 
\centerline{ \bf WITH A CHAOTIC NONLINEAR SYSTEM BY FEEDBACK OF STATES}

\vspace*{0.035truein}
\vspace*{0.37truein}
\vspace*{10pt} \centerline{\footnotesize G. SOLIS-PERALES$^1$\footnote{E-mail: solisp@fi.uady.mx} ,
H.C. ROSU$^2$\footnote{E-mail: hcr@ipicyt.edu.mx} , C. HERNANDEZ-ROSALES$^2$\footnote{E-mail: heros@ipicyt.edu.mx \hfill physics/0410031}  }
\vspace*{0.015truein}
\centerline{\footnotesize $^1$ Universidad Autonoma de Yucatan, Apdo Postal 150, Cordemex 97310 M\'erida, Yucat\'an, M\'exico}
\centerline{\footnotesize  $^2$ Departamento de Matem\'aticas Aplicadas
y Sistemas Computacionales, IPICYT, } \centerline{\footnotesize
Apdo. Postal 3-74 Tangamanga, 78231 San Luis Potos\'{\i}, Mexico }
\vspace*{0.225truein}


\vspace*{0.21truein} \abstracts{ {\bf Abstract.} A synchronization procedure of the generalized type in the sense of 
Rulkov {\em et al} [Phys. Rev. E 51, 980 (1995)]
is used to impose a nonlinear Malasoma chaotic motion 
on the Frenet-Serret system of vectors in the differential geometry of space curves. This could have applications to the mesoscopic motion of biological filaments. 
}{}{}


\textlineskip                  
\vspace*{12pt}                 

\vspace*{1pt}\textlineskip  
\vspace*{-0.5pt}
\noindent


\noindent

\newpage




\bigskip

\noindent
The Frenet-Serret (FS) vector set is in much use whenever one focuses on the kinematical properties of space curves. The evolution in time of the FS triad is 
one of the most used descriptions of the motion of tubular structures such as stiff (hard to bend) polymers \cite{Kamien02}. Many biological polymers including the 
DNA helical molecule are stiff and their movement is of fundamental interest. At the mesoscopic level many sources of noise and chaotic behaviour 
affect in a substantial way the motion of the biological polymers. In general, one can think that a synchronization between the motion of the polymers and the 
chaotic (or noisy) sources could be achieved in a natural way through some control signal. We illustrate this idea employing a  
generalized synchronization procedure based on the theory of nonlinear control by which we generate a chaotic dynamics of the FS evolution equations
\begin{eqnarray*} 
\dot{\vec{T}} & = &\kappa \cdot \vec{N}\\
\dot{\vec{N}} & = &\tau \cdot \vec{B}-\kappa \cdot \vec{T}\\
\dot{\vec{B}} & = &-\tau \cdot \vec{N}~.\\
\end{eqnarray*}
With this goal in mind, one should first write the FS system in the form
\begin{equation}\label{form}
\dot{x}=Ax+EU \qquad \qquad y=Cx~,
\end{equation}
where $x\in R^3$ is the vector having as components the tangent $T$, normal $N$ and the binormal $B$, whereas $A$ is the transfer matrix,
$E$ is an initial vector that determines the channel where the control signal is applied, and $C=(c_1,c_2,c_3)$ determines the measured signal of the FS system.

In this way, the main objective, from the synchronization standpoint, is to force the states of the slave system, which is the FS system, to follow the trajectories of 
the master system that, in general, presents a chaotic behaviour. This is achieved by applying, in a well-defined way, a signal given by $U=\Phi(x)$ and rewriting the 
FS system in the form given in Eq.~(\ref{form})
\begin{equation} \label{matrix}
\left( \begin{array}{ccc}
\dot{\vec{T}}\\
\dot{\vec{N}}\\
\dot{\vec{B}}
\end{array} \right )
=
\left(\begin{array}{ccc}
 0 & \kappa & 0 \\
-\kappa & 0 & \tau\\
0 & -\tau & 0\end{array} \right )\left( \begin{array}{ccc}
\vec{T}\\
\vec{N}\\
\vec{B}
\end{array}\right)+ \left( \begin{array}{ccc}
\vec{0}\\
\vec{0}\\
\vec{1}
\end{array}\right)\Phi~
\end{equation}
$$
y=(c_1 \, \, c_2 \,\,c_3)\left( \begin{array}{ccc}
\vec{T}\\
\vec{N}\\
\vec{B}
\end{array}\right)
$$
where, $c_1=1$, $c_2=0$, $c_3=1$.
If we choose now $B=(0\,\, 0\,\, 0)^T$ as the input vector, one gets the open loop FR dynamics. This type of dynamics is shown in Fig.~1 for $\kappa =1$,
$\tau =0.9$, and initial conditions given by $T_0=0.0024$, $N_0=0.0026$, and $B_0=0.0039$, where the states of the system are given by the derivatives of 
the tangent, normal, and binormal unit vectors, respectively, whereas $\kappa$ and $\tau$ are the curvature and torsion scalar invariants.

On the other hand, a chaotic oscilator is a dynamic system whose evolution is difficult to predict. In general, its main feature is the sensibility to the initial conditions
and the variations of the parameters. Thus, its long-term behaviour is hard to estimate.
In the following, we will use one of the simplest chaotic oscillator systems that has been introduced by Malasoma \cite{M02}
\begin{eqnarray} \nonumber
\dot{X_1} & = &X_2\\ 
\dot{X_2} & = &X_3\\ \nonumber
\dot{X_3} & = &-\alpha X_3-X_1+X_1X_2~,
\end{eqnarray} 
where $\alpha$ is the bifurcation parameter.
In matrix form, we get
\begin{eqnarray} \nonumber
\left( \begin{array}{ccc}
\dot{X_1}\\
\dot{X_2}\\
\dot{X_3}
\end{array}\right)
&=&
\left(\begin{array}{ccc}
 0 & 1 & 0 \\
0 & 0 & 1\\
-1& X_1 & -\alpha \end{array} \right )\left( \begin{array}{ccc}
X_1\\
X_2\\
X_3
\end{array}\right)\\
y&=&(1\,\,0\,\,0)x~.
\end{eqnarray}
This system exhibits chaotic behaviour for $2.0168 < \alpha < 2.0577$. The chaotic evolution is shown in Fig.~2 for $\alpha =2.025$ and initial conditions 
$X_1=0.0022$, $X_2=0.0024$, $X_3=0.0039$. Having the two systems in the matrix form, we choose the Malasoma one as the master system and the FS 
system as the slave one, i.e., the Malasoma dynamics will be imposed to the FS motion through the signal $U=\Phi(x)$. 
That means that a nonlinear dynamics is forced upon the FS system leading to its chaotic behaviour. 

To get the chaotic FS system one should achieve the synchronization between the master (of subindex $M$ in the following) and the slave systems
(of subindex $S$ in the following) . For this, one defines a third system,
which refers to the synchronization error given by the difference in the dynamics of the two systems, i.e.,  
\begin{eqnarray} \nonumber
\dot{e_1} & = &\kappa e_2+\lambda _1(X_M)\\ 
\dot{e_2} & = &\tau e_3-\kappa e_1+\lambda _2 (X_M)\\ \nonumber\label{error}
\dot{e_3} & = &-\tau e_2+\lambda _3(X_M)-U~,
\end{eqnarray}
where
\begin{eqnarray} \nonumber
\lambda _1 & = &X_{2M}(1-\kappa)\\ 
\lambda _2 & = &X_3(1-\tau)+\kappa X_{1M}\\ \nonumber \label{lambda}
\lambda _3 & = &-\alpha X_{3M}+X_{1M}(X_{2M}-1)+\tau X_{2M}~.
\end{eqnarray} 
The function $U=\Phi(e)$ gives the control action that leads to the synchronization of the two systems. 

Once (5)
defined, one simply choose $y=C=(1\,\, 0\,\, 0)e$ (or $y=e_1$) as the output of the error system.
In the synchronization approach, one writes $y=h(e)=e_1=X_{1M}-X_{1S}$, and consequently the error system (5) 
can be written in the general form
\begin{equation}\label{genform}
\dot{X}=f(e)+g(e)U~.
\end{equation}
The error system (\ref{genform}) should be stabilyzed at the origin or in an arbitrary small neighbourhood of it. More details on the synchronization conditions are
provided in the papers \cite{F97,SP03} that we employ to obtain the control function \cite{I89}
\begin{equation}\label{fcontrol}
U=\frac{-1}{\gamma}(\beta +\delta)~,
\end{equation}  
where $\gamma$ and $\delta$ are real-valued functions obtained by means of Lie derivatives of $h(e)$ as follows
\begin{equation}\label{gd}
\gamma = L_{g} L_{f}^{\rho -1}h(e)~, \qquad \qquad \beta = L_{f}^{\rho}h(e)~,
\end{equation}
where $\rho$ is a positive integer that determines the so-called relative degree of the system (see \cite{I89}).
On the other hand, the desired dynamics, i.e., directed towards the origin, is dictated by
\begin{equation}\label{delta1}
\delta = K_{P1}e_1+ K_{P2}e_2+ K_{P3}e_3~.
\end{equation}
Thus, performing the Lie derivatives and regrouping the terms, we obtain the function $U$ of the form
\begin{equation}\label{Uform}
U=\frac{-1}{\kappa \tau}\Big[\left( -\kappa ^2 \dot{e_1}+\kappa \tau \dot{e_3}\right)+K_{P1}e_1+ K_{P2}e_2+ K_{P3}e_3\Big]~.
\end{equation}
Using the change of variables $e_i=X_{iM}-X_{iFS}$, where the latter vector is the column vector formed by the triad $T$, $N$, and $B$, ($i=1,2,3$), the control can be written as a function of the states of the two systems 
\begin{equation}\label{delta2}
U=\Phi(x)=\frac{-1}{\kappa \tau}\Bigg[-\kappa ^2\left(X_2-\kappa N\right)-\kappa \tau \Big[\left(-\alpha X_3-X_1+X_1X_2\right) +(\tau N)\Big]+ \delta \Bigg]~,
\end{equation}
where $\delta=K_{P1}(X_1-T)+K_{P2}(X_2-N)+K_{P3}(X_3-B)$.
Notice that $\gamma = L_gL^{2}_{f} h(x)=\kappa \tau$ is a nonzero constant. Therefore the control signal is defined for any $T$,$N$,$B$, $X_1$,$X_2$, and $X_3$.
In addition, one should choose the values of the constants $K_{Pi}$ in such a way that the differences in $\delta$ go to zero. 
Applying the dynamics generated by (\ref{delta2}) leads to the synchronization matrix 
\begin{equation}\label{matrixs}
X_{FS}=
\left(\begin{array}{ccc}
 X_{1M} & 0 & 0 \\
 0 & X_{2M} & 0\\
\kappa X_{1M}& 0& X_{3M}\end{array} \right )~.
\end{equation}
From this synchronization matrix one can see that the first two states of both systems are synchronized. However,
the state $B=X_{3M}+\kappa X_{1M}$ is extended by the term $\kappa X_{1M}$, i.e., the state $B$ is the sum of two states of the Malasoma oscillator;
since the latter is chaotic, one concludes that the state $B$ is also chaotic.

We display the phase locking between the corresponding phases of the two oscillators in
Figs.~3  and 4 where the phase locking of the states $T$ and $X_1$ and $N$ and $X_2$, respectively, shows that the two pairs of states are synchronized.
In Fig.~5, we see that the $B$ and $X_3$ states are not synchronized. Thus, following the terminology of \cite{R95}, we are in the situation of a generalized
synchronization. 
In Fig.~6, the two already synchronized systems are shown in the three dimensional space. One can notice that the FS system 
is `above' the Malasoma oscillator, and that the two systems are in a chaotic phase.
Finally, in Fig.~7, the control signal used to achieve the generalized synchronization of this paper is displayed.   

In summary, we have shown here in a concrete way how the simple chaotic dynamics of Malasoma type can be imposed to the linear Frenet-Serret evolution of
space curves.


%


\bigskip

\begin{figure}[htb]   
\centerline{
\includegraphics[scale=1]{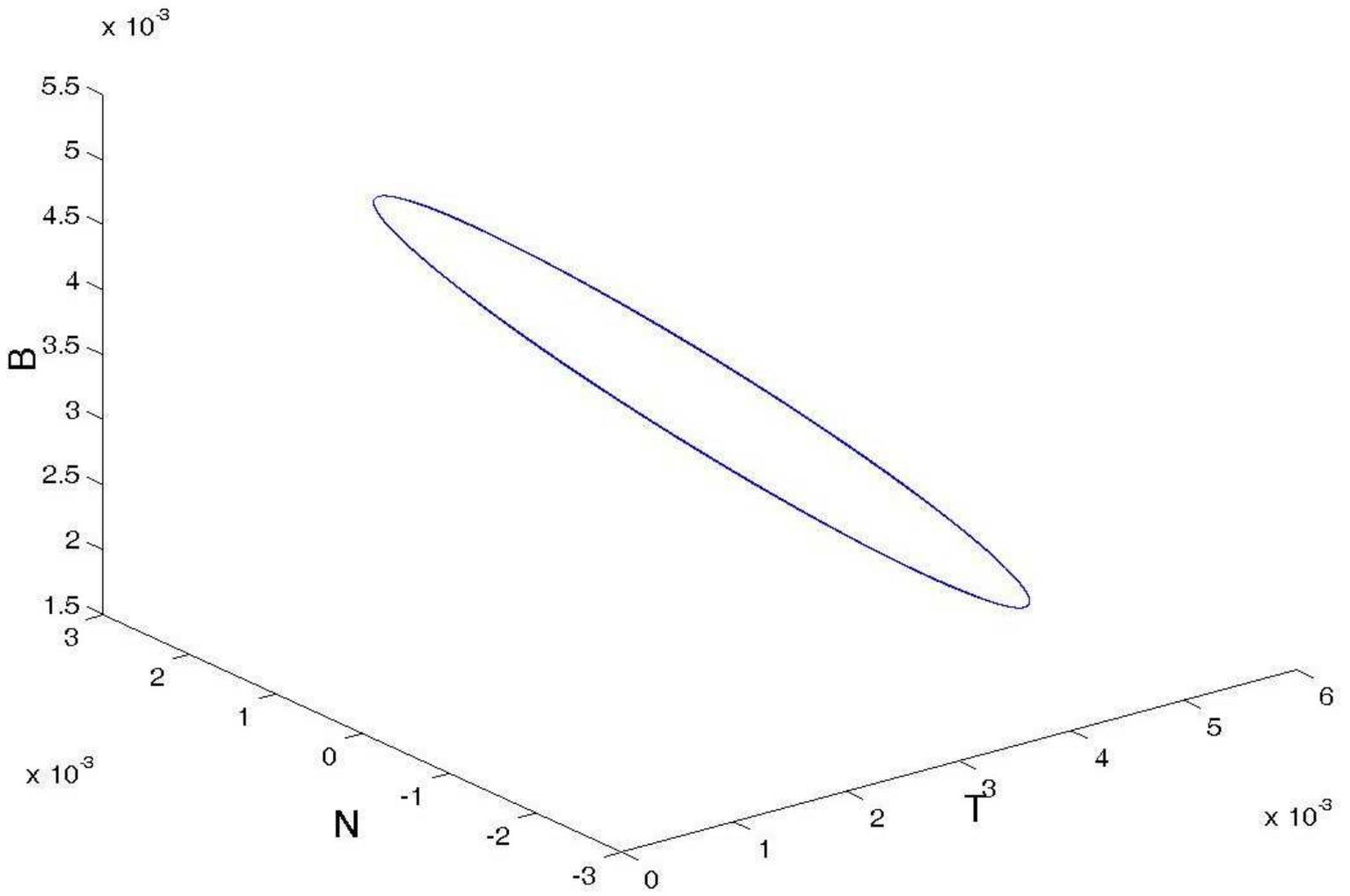}}
\caption{Three dimensional dynamics of the Frenet-Serret system with $\kappa=1$ and $\tau = 0.9$.} \label{fig1fs}
\end{figure}

\begin{figure}[htb] 
\centerline{
\includegraphics[scale=1]{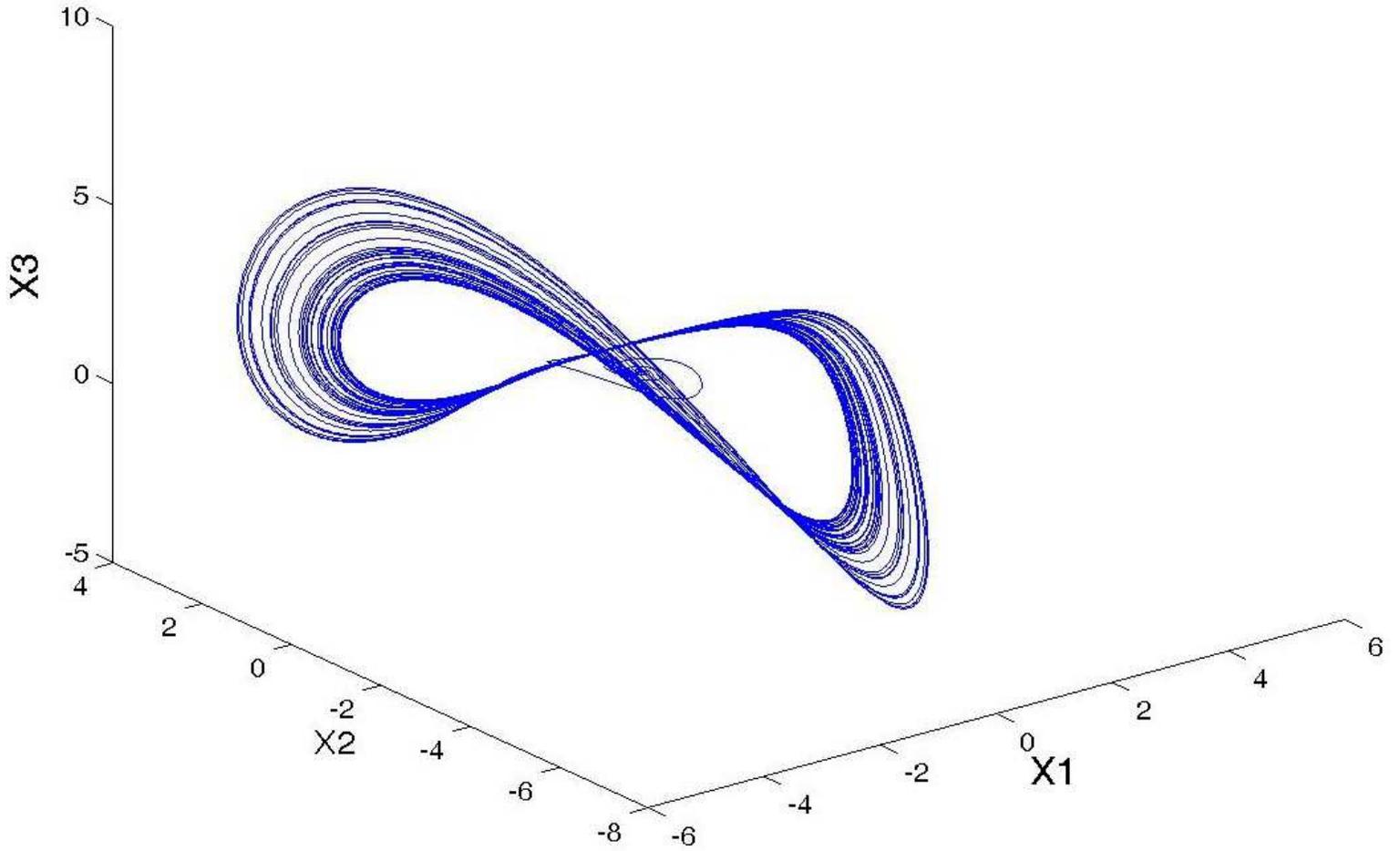}}
\caption{Three dimensional dynamics of the Malasoma system with $\alpha =2.025$.} \label{fig2fs}
\end{figure}

\begin{figure}[htb] 
\centerline{
\includegraphics[scale=1]{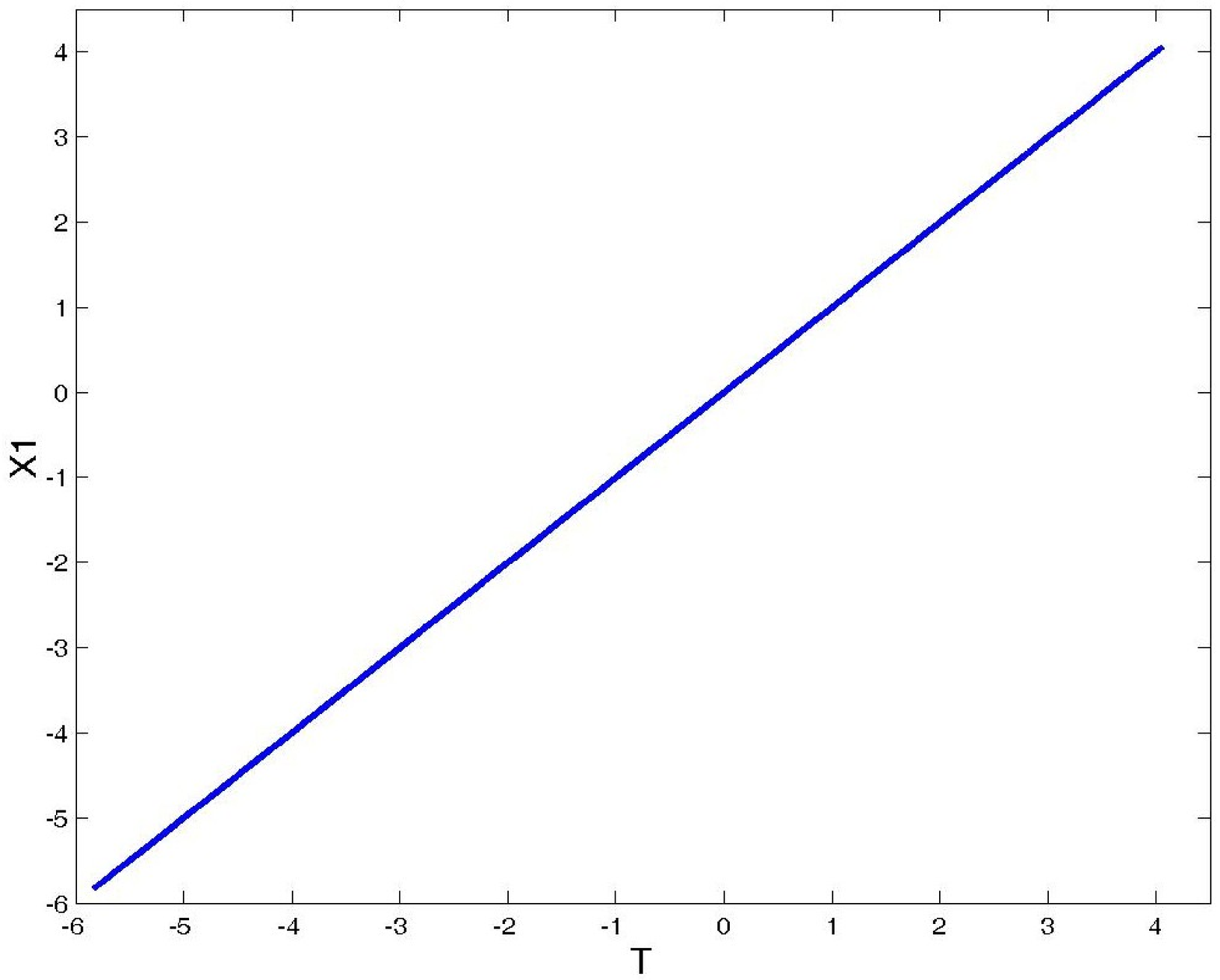}}
\caption{Phase locking of the $T$ states of the FS system to the $X_1$ states of the Malasoma system.} \label{fig3fs}  
\end{figure}

\begin{figure}[htb] 
\centerline{
\includegraphics[scale=1]{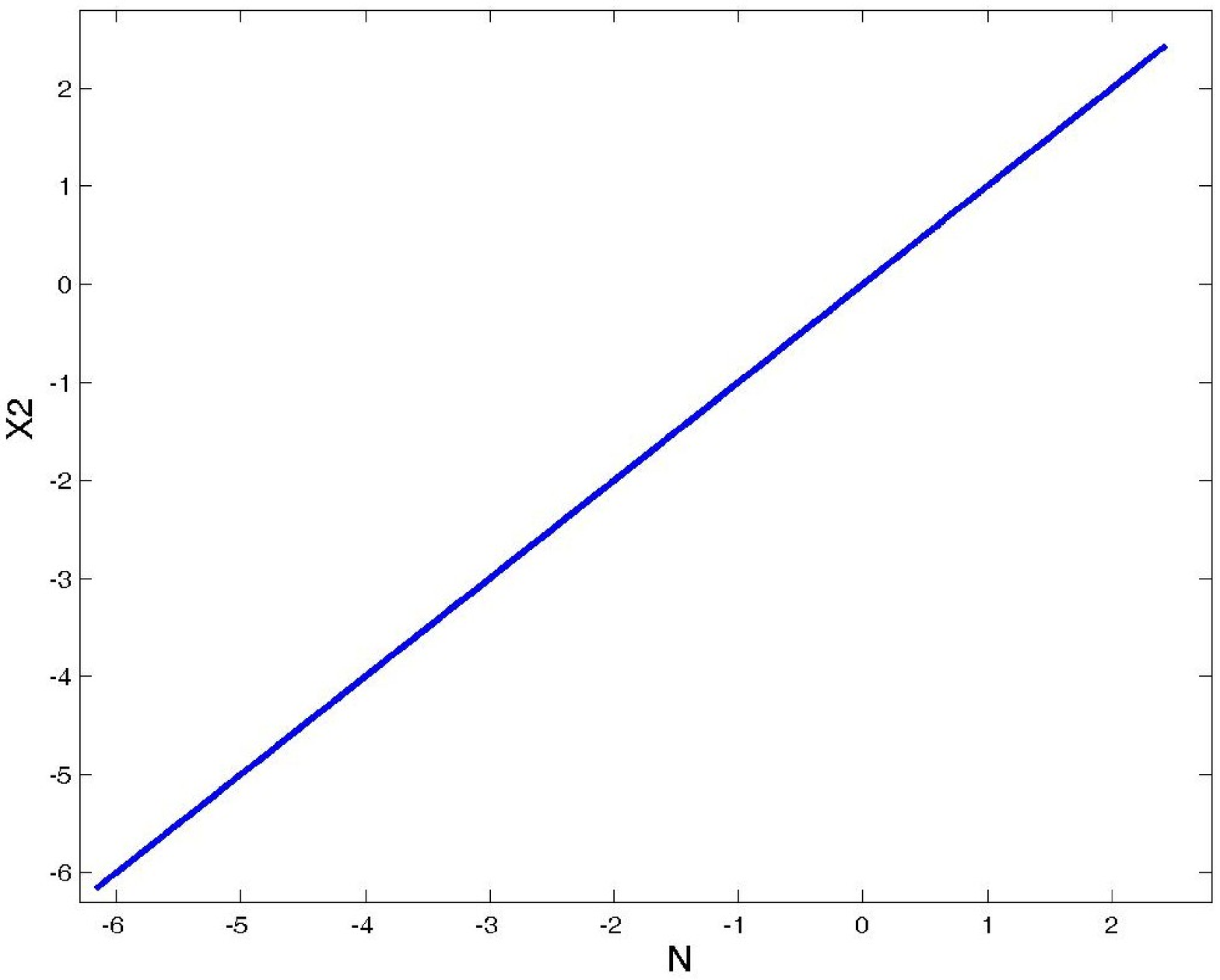}}
\caption{Phase locking of the $N$ states of the FS system to the $X_2$ states of the Malasoma system.} \label{fig4fs}
\end{figure}

\begin{figure}[htb]  
\centerline{
\includegraphics[scale=1]{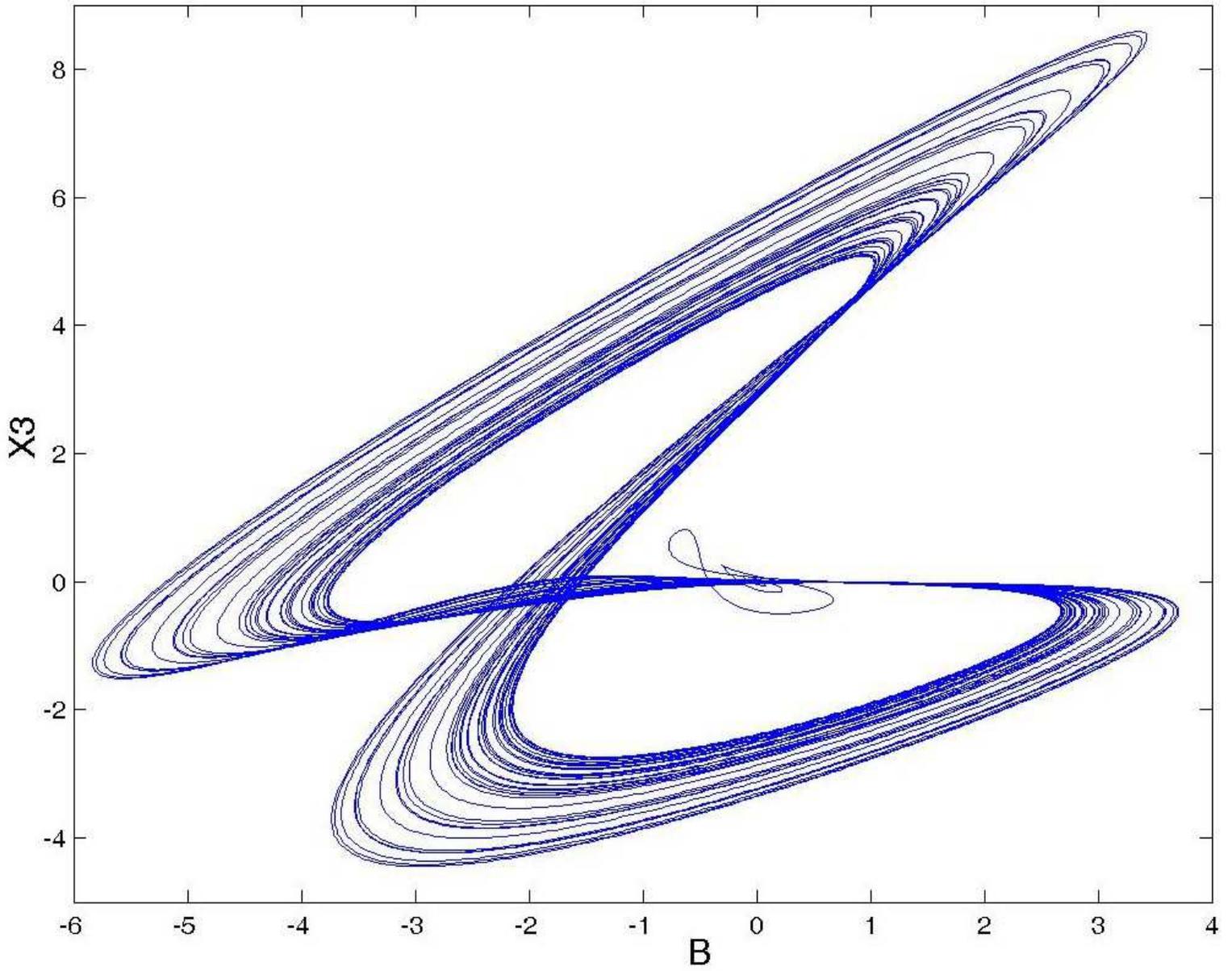}}
\caption{Phase behaviour of the states $B$ and $X_3$ of the FS system and Malasoma system, respectively.} \label{fig5fs}
\end{figure}
%

%

\begin{figure}[htb]  
\centerline{
\includegraphics[scale=1]{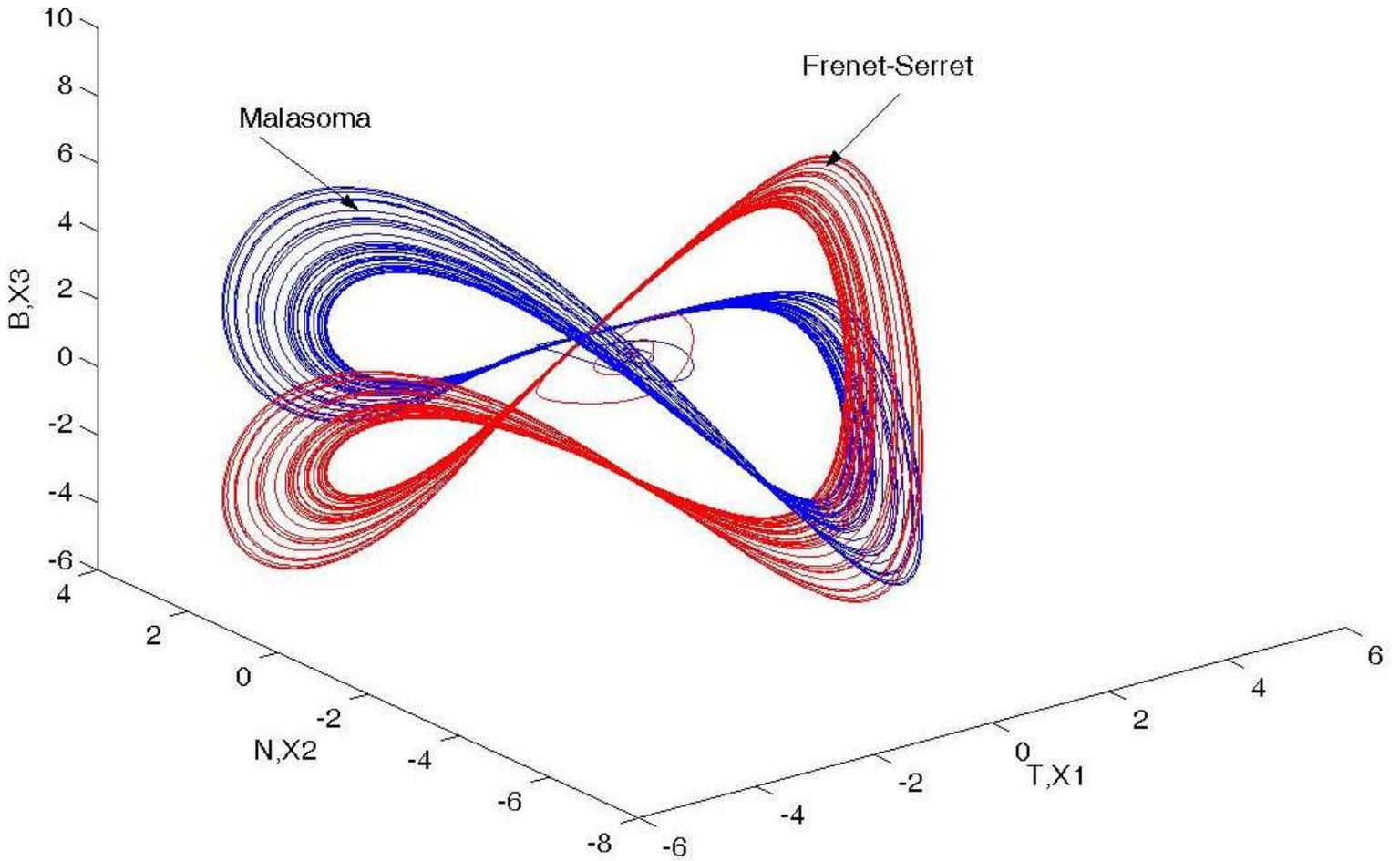}}
\caption{Plot of the two chaotic attractors once they are synchronized in the generalized form.} \label{fig6fs}
\end{figure}

\begin{figure}[htb]  
\centerline{
\includegraphics[scale=1]{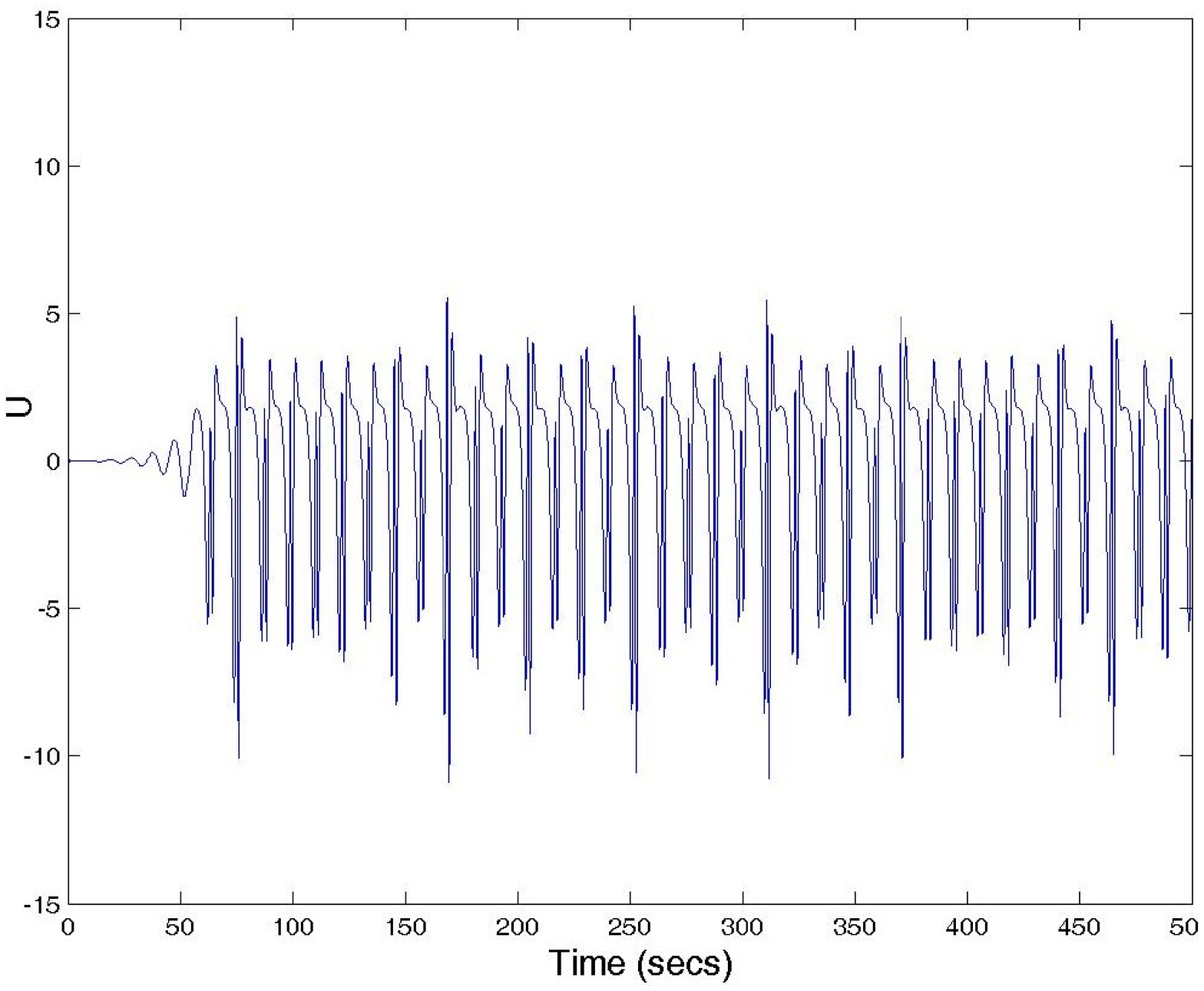}}
\caption{The control signal ${\rm U}$ imposed to the Frenet-Serret system in order to display Malasoma chaos.} \label{fig7ho}
\end{figure}


\begin{thebibliography} {000}


\bibitem{Kamien02} Kamien, R.D.,
``The geometry of soft materials: a primer",
Rev. Mod. Phys. {\bf 74}, 953 (2002).

\bibitem{M02} Malasoma, J.-M.,
``A new class of minimal chaotic flows",
Phys. Lett. A {\bf 305}, 52 (2002).

\bibitem{F97} Femat, R. and Alvarez-Ramirez, J.,
``Synchronization of a class of strictly different oscillators",
Phys. Lett. A {\bf 236}, 307 (1997).

\bibitem{SP03} G. Solis-Perales, G., Ayala, V., Kliemann, W. and Femat, R.,
``On the synchronizability of chaotic systems: A geometric approach",
Chaos {\bf 13}, 495 (2003).

\bibitem{I89} Isidori, A.,
``Nonlinear Control Systems''
(Springer, Berlin 1989).


\bibitem{R95} Rulkov, N.F., Sushchik, M.M., Tsimring, L.S. and Abarbanel, H.D., 
``Generalized synchronization of chaos in directionally coupled chaotic systems",
Phys. Rev. E {\bf 51}, 980 (1995).

\end{thebibliography}
\end{document}